# Testing theoretical models of magnetic damping using an air track

Ana Vidaurre[*], Jaime Riera, Juan A. Monsoriu, and Marcos H. Giménez

Department of Applied Physics, Polytechnic University of Valencia, E-46022 Valencia,

Spain

**ABSTRACT**

Magnetic braking is a long-established application of Lenz's law. A rigorous analysis of the laws governing this problem involves solving Maxwell's equations in a time-dependent situation. Approximate models have been developed to describe different experiences related to this phenomenon. In this paper we present a new method for the analysis of the magnetic braking using a magnet fixed to the glider of an air track. The forces acting on the glider, a result of the eddy currents, can be easily observed and measured. As a consequence of the air track inclination, the glider accelerates at the beginning, although it asymptotically tends towards a uniform rectilinear movement characterized by a terminal speed. This speed depends on the interaction between the magnetic field and the conductivity properties of the air track. Compared with previous related approaches, in our experimental setup the magnet fixed to the glider produces a magnetic braking force which acts continuously, rather than over a short period of time. The experimental results satisfactorily concur with the theoretical models adapted to this configuration.

---

[*] E-mail: *vidaurre@fis.upv.es*

# I. INTRODUCTION

When a conductor material is under the effect of a changing magnetic flux, eddy currents are induced in the conductor. This change in the flux can be produced either because the conductor is moving in a region where there is a magnetic flux or, similarly, because the magnet is moving. The action of the magnetic field on the induced currents produces a braking force. A rigorous analysis of the laws governing the problem entails the solving of the Maxwell equations in a time-dependent situation. This depends on the problem geometry and it is usually difficult to solve.

Widerick et al.[1] present a very simple model for the calculation of the magnetic drag force on a moving metal disc in the air gap between the rectangular-shaped pole pieces of an electromagnet. Likewise, Cadwell[2] analyzes the effect of magnetic damping on an aluminium plate moving on a horizontal air track as it passes between the poles of a horseshoe magnet. In both cases, it is assumed that the induced current in the "shadow" of the magnet is uniform and equal to $\mathbf{J} = \sigma(\mathbf{v} \times \mathbf{B})$. Both papers present a simple model and fail when trying to explain the influence of the magnet size or its position with respect to the motion direction. Heald[3] replaces the simplification of considering the eddy current density as a constant with a more realistic approach which also takes into account the contribution of the electric field generated by the charge separation: $\mathbf{J} = \sigma(\mathbf{E} + \mathbf{v} \times \mathbf{B})$. If we select the reference system in such a way that OX is the direction of the motion, then the Coulomb sources of $\mathbf{E}$ are the surface charges, within the conducting sheet, on planes perpendicular to the OX axis. Therefore the magnetic braking force depends on the aspect ratio, A=Length/width, of the magnet footprint. Marcuso et al.[4,5] apply a method of successive approximations to solve the Maxwell equations, in the case of a conducting disk rotating in the externally applied non-uniform magnetic field. The experimental results satisfactorily concur with the theory except for the area near the disk border. Aguirregabiria et al.[6] study the same problem in a quasi-static approximation, emphasizing the role played by the charge distributions induced in the disk. In this paper, cases of both infinite and finite radii are considered in order to analyse the border effects. Similarly, Lee and Park[7,8] present the model and experimental results for a rotating disk with a rectangular-shaped electromagnet. They consider the boundary conditions of the rotating disk by using the mapping and image method techniques. Salzman et al.[9], in a pedagogical manner, reveal the solution to the problem of a very large plane conducting sheet passing between circular magnet poles. They emphasize the importance of considering the induced electric field, as is pointed out by Gauthier[10]. Related studies concerning damping forces due to eddy currents on oscillating systems[11,12], or those forces present when a magnet moves through a pipe[13,14], serve to illustrate different aspects of the same main problem. In all the cited papers the conductor passes between the poles of a fixed magnet. Subsequently, the effects can only be measured over a short period of time which can be repeated periodically in the case where the conductor is a rotating disk.

In this paper we present a new experimental setup where the magnet is fixed to a glider sliding on an air track. From the physical point of view, the movement of the magnet close to a resting conductor is equivalent to the movement of the conductor close to a fixed magnet. The materials needed (air track, glider, magnet...) are commonplace laboratory materials for first year undergraduate Physics courses. With this configuration, the magnetic damping force does not only act over a short period of time, but also acts continuously during the whole movement. Moreover, the influence of the aspect ratio of the rectangular magnet footprint can be easily analysed by rotating the magnet orientation on the glider. However, the one drawback of this device is that the magnetic field is not uniform and this makes the theoretical problem more complicated.

In previous studies, experimental results have been achieved using different methods: a commercial Pasco motion sensor[1,2], a photoresistor connected to a microcomputer[5], by taking measurements of the braking torque through reading the output voltage of the load cell[7,8]. Another method has been to use an oscilloscope[14] and computer to record the voltage pulse[11-13]. We propose to take measurements of the position as a function of time by means of digital image capture which has proved to be an effective method[15,16,17]. This method can be automated by using image recognition techniques[18,19]. We used standard linear correlation[20] as a basis for the detection technique. The experimental results are compared with the theoretical motion equations solved by applying the theoretical model. Furthermore, the experimental setup has been designed to show the relevance of the induced electric field.

## II. THEORETICAL ANALYSIS

According to Faraday's law, when a magnet fixed to the glider is moving on an air track, the changing magnetic flux through the aluminium track produces an electromotive force equal to the time rate of change in the magnetic flux given by

$$\varepsilon = -\frac{d\phi}{dt} = -\frac{d}{dt}\int \mathbf{B}\cdot d\mathbf{A} \ . \tag{1}$$

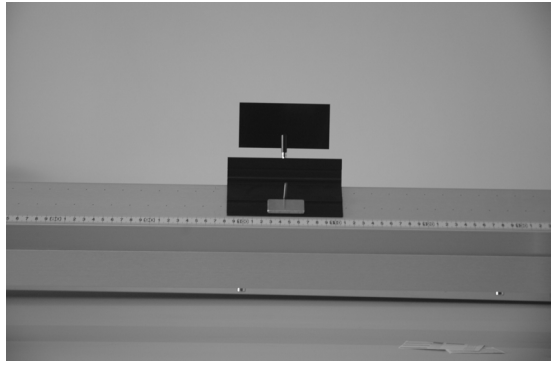

Figure 1. Experimental setup: a magnet is fixed to a glider sliding on an air track.

Figure 1 shows the experimental design. In an initial approximation it can be considered that the magnetic field is perpendicular to the conducting plane: $\mathbf{B} = B\hat{\mathbf{z}}$. The reference system moves with the glider, so that the magnet is at rest and the electromotive force produces a current density inside the conducting air track that is represented by

$$\mathbf{J} = \sigma(\mathbf{E} + \mathbf{v} \times \mathbf{B}), \qquad (2)$$

where $\sigma$ is the conductivity of the air track material, $\mathbf{E}$ is the electrostatic field of Coulomb charge induced within the conductor, $\mathbf{v} = v\hat{\mathbf{x}}$ is the velocity of the conducting sheet relative to the glider, and $\mathbf{B}$ is the magnetic field measured at rest. Some authors[1,2] only consider the second term of Eq. (2). However, the magnitude of the first term can be as large as the second one and, since these take opposite directions, the net current density could even be zero. This would occur in the limit case of a very long and thin magnet (as applied to the motion direction).

From the microscopic point of view[21], the magnetic field, acting on the moving charges of the conductor produces a force $\mathbf{F} = q(\mathbf{v} \times \mathbf{B})$ in these. This force causes the charges of different signs to separate. This separation of charges will produce an electric field pointing upward that will tend to decrease the total force on the given charge moving in the conductor material (see Fig. 2). As can be seen in Fig. 3, the magnetic field in the shadow region is not uniform. As such, the exact solution would be arrived at through solving the Maxwell equations. However, the calculation of the electric field can be simplified when taking a uniform magnetic field $B_{avg}$, which is the average magnetic field obtained from the experimental data measured at positions on a hypothetical three-dimensional grid. To obtain this electric field, let us consider the surface charge density as the product of the polarization and the unit outward normal vector[3,8] $\rho_s = \varepsilon_0 (\mathbf{v} \times \mathbf{B}) \cdot \hat{\mathbf{u}}$. Hence, we will have $\rho_{s+} = -\varepsilon_0 v B_{avg}$ and $\rho_{s-} = +\varepsilon_0 v B_{avg}$ at the planes perpendicular to the OY axis and at $y = +a/2$ and $y = -a/2$, respectively.

The electric field intensity $\mathbf{E}$ is represented by $\mathbf{E} = E_x \hat{\mathbf{x}} + E_y \hat{\mathbf{y}}$, where $E_x$ and $E_y$ are obtained using Coulomb's law. As pointed out in Refs. 3,7 and 8, although the thickness of the conductor (in the z direction in our reference system) is small compared with the a, b dimensions, the net result is such that the interior electric field of the capacitor-like surface charges $\rho_{s\pm} = \mp \varepsilon_0 v B_{avg}$ at $y = \pm a/2$ extend indefinitely in the z direction, and from $x = -b/2$ to $x = +b/2$ in the x direction. On this charged surface, the linear infinitesimal element of length $d\xi$ in the y direction and infinite length along the OZ direction can be assumed. If $\rho_{\ell+} = -\varepsilon_0 v B_{avg} d\xi$ and $\rho_{\ell-} = +\varepsilon_0 v B_{avg} d\xi$ are defined as the line charge densities at the points $(\xi, a/2)$ and $(\xi, -a/2)$ respectively, the electric field intensity at the point P(x,y,0) takes the form, from Coulomb's law[21],

$$\mathbf{E} = \int \frac{\rho_{\ell+}}{2\pi\varepsilon_0} \frac{\hat{\mathbf{r}}_+}{r_+} + \int \frac{\rho_{\ell-}}{2\pi\varepsilon_0} \frac{\hat{\mathbf{r}}_-}{r_-} = \int \frac{\rho_{\ell+}}{2\pi\varepsilon_0} \frac{\mathbf{r}_+}{r_+^2} + \int \frac{\rho_{\ell-}}{2\pi\varepsilon_0} \frac{\mathbf{r}_-}{r_-^2}, \qquad (3)$$

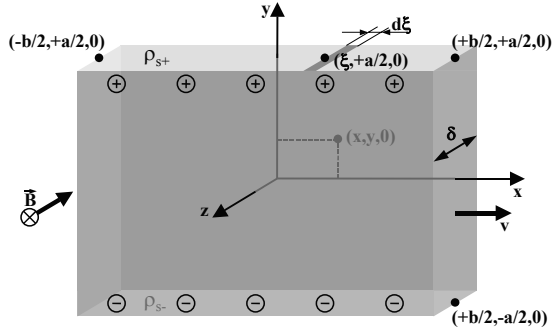

Figure 2. Charge distribution and electric field induced in the conductor.

where $\mathbf{r}_\pm$ has coordinates $\mathbf{r}_\pm = (x-\xi, y \mp \frac{a}{2})$ as shown in Fig. 2. By performing the corresponding integration, the following is obtained

$$E_x = \frac{vB_{avg}}{4\pi}\left[\ln\frac{(2x-b)^2+(2y+a)^2}{(2x+b)^2+(2y+a)^2} - \ln\frac{(2x-b)^2+(2y-a)^2}{(2x+b)^2+(2y-a)^2}\right], \quad (4a)$$

$$E_y = \frac{vB_{avg}}{2\pi}\left[\tan^{-1}\frac{2x-b}{2y+a} - \tan^{-1}\frac{2x+b}{2y+a} + \tan^{-1}\frac{2x+b}{2y-a} - \tan^{-1}\frac{2x-b}{2y-a}\right]. \quad (4b)$$

The magnetic braking force is the component opposite to the velocity of the Lorentz force:

$$F = \int (\mathbf{J} \times \mathbf{B})_x \, d\tau = -\sigma \int (E_y - vB_{avg}) B \, d\tau. \quad (5)$$

In this expression the current density and the magnetic field are functions of the position and the total force is performed by integration. As the vertical component of the electric field is proportional to the velocity and to the average magnetic field, this can be written in the form: $E_y = vB_{avg}f(A, x, y)$, where A=b/a is the aspect ratio of the magnet footprint. The corresponding braking force

$$F = -v\sigma B_{avg} \int_{-b/2}^{b/2} \int_{-a/2}^{a/2} \int_{-\delta/2}^{\delta/2} B(1 - f(A,x,y)) dx dy dz =$$
$$= v\sigma B_{avg} \iiint_\tau B(1 - f(A,x,y)) d\tau \qquad (6)$$

may be rewritten as

$$F = -m\alpha v, \qquad (7)$$

where $\alpha$ is a coefficient that depends on the geometry of the magnet footprint as well as on the magnetic field

$$\alpha = \frac{\sigma B_{avg}}{m} \iiint_\tau B(1 - f(A,x,y)) d\tau. \qquad (8)$$

This result is similar to that obtained by Heald[2]. However, in that article the conductor is a rotating disk, but they consider the magnet in the OY axis and then the velocity in the OX direction. On the other hand, Lee and Park[7,8] take into account the size of the shadow magnet, and therefore the x and y components of the velocity, depending on the position. As a consequence they have four charged surfaces and the problem takes longer to solve, but not more complicated.

We have numerically performed the integration given by Eq. (8). For each one of the points on the three-dimensional grid where we have measured the magnetic field B, we have also calculated the function f(A,x,y) given by Eq. (4a). The result of the integrand of Eq. (8) has been multiplied by the volume element, and then we have added all the terms obtained in this manner.

## III. EXPERIMENT DESIGN

To validate the theoretical expectations we have measured the effect of the braking force on the movement of a glider that has a magnet fixed to it. The system glider-magnet moves on an inclined frictionless air track. Subsequently, the acceleration of the system depends on the tangential component of the weight and the braking force given by Eq. (7).

The air track is a commercial PASCO Scientific® that has a length of 2 m, and is made of aluminium with a thickness of 3 mm (see Fig. 1). The electric conductivity is 3.1 $10^7$ $(\Omega m)^{-1}$ (supplied by the manufacturer). The glider has a mass of 190 g and the magnet is fixed to it in such a way that the magnetic field lines pass through the glider (2 mm thick) and arrive at the conducting sheet in a perpendicular manner. The air track forms an angle of approximately 0.8º with the horizontal.

The commercial magnet, made of NdFeB, has a parallelepipedic shape of dimensions $4\times 2\times 0.5$ cm$^3$ and a mass of 25 g. The magnetic poles are on the 4x2 cm$^2$ surfaces. The resultant magnetic field in the conducting sheet (the air track) is not homogeneous. It has been measured by means of a gauss-meter - FW-Bell-4048 at 66 points of the magnet shadow at 3 different planes parallel to the magnet surface, 22 points in each plane. These planes are located at a distance of 2 mm from the magnet surface (this is the distance at which the outer surface of the air track is located), at 3.5 mm (in the middle of the air track) and, at 5 mm (the inner part of the air track). In Fig. 3, we can see a three dimensional representation of the magnetic field as it has been measured. The average magnetic field has been obtained through the expression

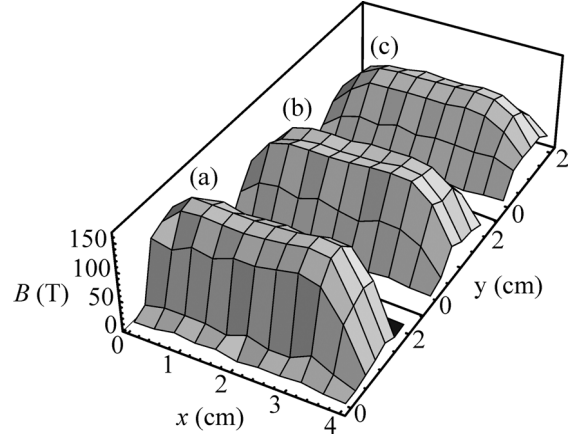

Figure 3. Magnetic field at three different distances from the magnet pole: (a) 0 mm, (b) 3.5 mm, and (b) 5 mm.

$$B_{avg} = \frac{1}{\tau}\sum B_i \Delta\tau_i. \qquad (9)$$

where $\tau$ is the total volume of the region of the conducting sheet, when some appreciable magnetic field is present; $\Delta\tau_i$ represents the volume of each element by which we have divided this region. The value of the force given by Eq. (7) depends on $\alpha$ Eq. (8), which can be numerically obtained using

$$\alpha = \frac{\sigma B_{avg}}{m}\sum B_i (1 - f(A,x,y))\Delta\tau_i, \qquad (10)$$

with m the total mass (glider, magnet and the counterweight placed to maintain a balanced glider mass).

The measurements of the glider position were obtained by means of the video-analysis technique[15-19]. The digital camera used in the experiments was a Panasonic NV-DS15EG, with an exposure time of 1/750 s and with a rate of 25 frames/s, providing a time resolution of 0.04 s. The camera was placed with its axis perpendicular to the movement direction, at a distance of 1.5 m. The video system was PAL (phase alternation by line) which can produce 720x576 pixel images. As most of the image recognition techniques require the use of matrices, the dimensions of which are integer powers of 2, we have taken 512x512 pixel windows for the information analysis.

IV. RESULTS AND DISCUSSION

In order to test our theoretical model, we have performed experimental measurements of the braking magnetic force by looking at its effect on the movement of the glider. Let us consider the glider sliding without friction on an inclined plane that makes an angle θ with the horizontal under the effect of the braking force given by Eq. (7). If we take the OX axis along the movement direction, the motion equation, given by Newton's second law of motion, is

$$m\frac{d^2x}{dt^2} = mg\sin\theta - m\alpha\frac{dx}{dt} \tag{11}$$

the solution of which can be written in the following way

$$x = x_0 + \frac{1}{\alpha}(v_0 - v_T)[1 - \exp(-\alpha t)] + v_T t \tag{12}$$

where $x_0$ is the initial position, $v_0$ is the initial velocity, and $v_T$ is the terminal velocity given by

$$v_T = \frac{g\sin\theta}{\alpha} \tag{13}$$

with g being the gravity acceleration.

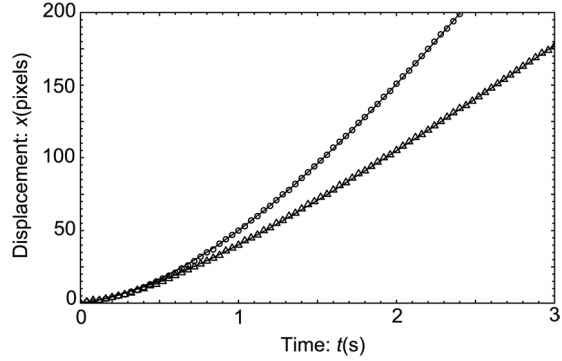

Figure 4. Representation of the displacement (Δx) vs. time period for two different aspect ratios of the magnet footprint, A=2 (circles) and A=0.5 (triangles). In both cases, the average magnetic field is B=66 mT.

From the experimental results of the position as a function of time, the distance travelled Δx=x-x$_0$ was fitted to the theoretical expression

$$\Delta x = D[1 - \exp(-\alpha t)] + v_T t, \qquad (14)$$

where the fitting parameters are the dumping coefficient, $\alpha$, the terminal velocity, $v_{lim}$, and parameter D, which is related to the initial velocity, $v_o = v_{lim} + \alpha D$.

Figure 4 displays the experimental results and the corresponding numerical fits to Eq. (14) for two different aspect ratios of the magnet footprint: A=2 (horizontal) and A=0.5 (vertical), in both cases, this being the average magnetic field B=66 mT. The $\alpha$ parameter allows us to check the theoretical model, by comparing its calculated and experimental fitted results. As expected, the greater A is, the lower the value for the coefficient $\alpha$. Consequently, a lower dumping of the movement is obtained.

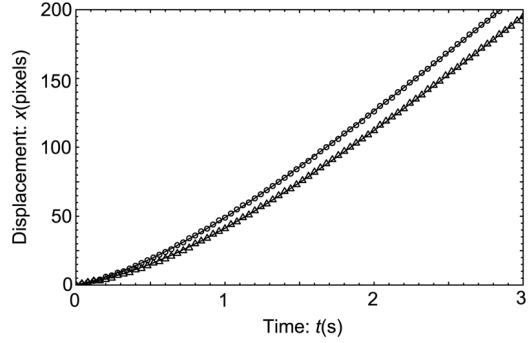

Figure 5. Representation of the displacement (Δx) vs. time period for two different aspect ratios of the magnet footprint, A=2 (circles) and A=4 (triangles). In both cases, the average magnetic field is B=78 mT.

In order to verify the importance of the electric field contribution to the drag force we have used a longer magnet with a footprint aspect ratio A=4. Figure 5 represents both the experimental and the fitted results together. For the sake of comparison, the corresponding results for A=2 are represented. In both cases the average magnetic field is B=78 mT. In this case, the shadow of the magnet is double the size of the other shadow represented. However, the expected values for the dumping coefficient are of the same order. This result is confirmed by the experiment.

In Table I a summary of the most important parameters entailed in this experience can be observed. The experimental parameters $\alpha_E$ and $v_T$, and their uncertainties, were obtained using the standard least-squares method. The comparison between the theoretical expectations and the experimental results allow us to state that the proposed theoretical model is a good match for this experience.

| **A** | **B** (mT) | $\alpha_T$ (s$^{-1}$) | $\alpha_E$ (s$^{-1}$) | **V$_T$** (pixels/s) |
|---|---|---|---|---|
| 2 | 66 | 0.60 | 0.64±0.02 | 156.8±1.6 |
| 0.5 | 66 | 1.12 | 1.11±0.03 | 76.65±0.18 |
| 2 | 77 | 0.84 | 1.03±0.03 | 92.2±0.3 |
| 4 | 77 | 0.87 | 0.87±0.02 | 92.1±0.4 |

Table 1. Numerical results for the four configurations analyzed in Figs. 4 and 5. The theoretical parameter $\alpha_T$ is obtained from Eq. (10). The experimental parameters, $\alpha_E$ and $v_T$, are obtained from the data fit given by Eq. (14).

## V. CONCLUSIONS.

As we have seen, our experimental setup allows students to investigate magnetic damping using the conventional materials found in laboratories for first year undergraduate Physics courses. Compared with previous related approaches, in our case a magnet is fixed to a glider that slides on an air track, producing a magnetic braking force that acts continuously. The results satisfactorily concur with the theoretical predictions. Furthermore, the relevance of the electric field induced in the conductor is demonstrated. The present study sheds new light on several physics experiences. In Particular, we are currently designing a new electromagnetically-damped, coupled oscillator system using this methodology.